\documentclass[12pt]{article}
\usepackage[utf8]{inputenc}
\usepackage[english]{babel}
\usepackage{amssymb,graphicx}
\usepackage{amsmath,amsfonts}
\usepackage{fullpage}
\def\e{{\rm e}}

\def\l{\left(}
\def\r{\right)}
\def\eps{\varepsilon}
\def\arcth{\mathrm{arcth}}
\newcommand{\be}{\begin{equation}}
\newcommand{\ee}{\end{equation}}

\renewcommand{\ln}{\mathop{\rm ln}\nolimits}
\newcommand{\Tr}{{\rm Tr}}

\author{
Petr Satunin\footnote{{\bf e-mail:} satunin@ms2.inr.ac.ru} \vspace{.2cm}\\
\normalsize\it Institute for Nuclear Research of the Russian Academy
of Sciences, \\  
\normalsize \it  60th October Anniversary Prospect, 7a, 117312  Moscow, Russia
} 
\title{
\vspace{-2cm} 
\begin{flushright}
{\normalsize INR-TH-2017-016}
\end{flushright}
\vspace{0.5cm} 
One-loop correction to the photon velocity in Lorentz-violating QED.}
\date{}
\begin{document}
\maketitle
\begin{abstract}
We calculate a finite momentum-dependent part of the photon polarization operator in a simple model of Lorentz-violating quantum electrodynamics nonperturbatively at all orders of Lorentz-violating  parameters. We sum one-particle reducible diagrams into the modified photon propagator, and determine the physical photon dispersion relation as the location of its pole. The photon dispersion relation, as well as its group velocity, acquires the one-loop momentum-dependent radiative correction.
% discuss dispersion properties of a photon. Photon group velocity becomes dependent on energy, a renormalization group explanation is provided.
We constrain the Lorentz-violating parameters for heavy charged fermions (muon, $\tau$-lepton, top-quark) from the photon timing observations.
\end{abstract}
\section{Introduction}

Small violation of Lorentz Invariance (LI) may take place in physics at low energies as a relic of some unknown ultraviolet theory, which includes quantum gravity. There are numerous approaches to quantum gravity, featuring Lorentz Invariance violation (LV), such as loop quantum gravity \cite{Girelli:2012ju}, noncommutative field theory \cite{Mocioiu:2000ip}, spacetime foam \cite{Ellis:2003if}, some approaches in string theory \cite{Kostelecky:1988zi}, Horava-Lifshitz gravity \cite{Horava:2009uw,Blas:2009qj} and others (see \cite{Mattingly:2005re,Liberati:2013xla,AmelinoCamelia:2008qg} for reviews). In accordance with these models, small but nonzero LV may appear in matter sector as well.

Lorentz Invariance violation in matter sector may be also considered phenomenologically. The most general framework describing LV in matter sector is called Standard Model Extension (SME) \cite{Colladay:1998fq}. The SME Lagrangian includes all possible operators of a given order, which are scalars under coordinate transformations. These operators are controlled by the coefficients which may be tested (and constrained) experimentally (see \cite{Kostelecky:2008ts}).

QFT methods in a theory without LI may be developed in analogy with corresponding LI theory. Such models can be quantized \cite{Kostelecky:2000mm}, and Feynman rules for perturbative calculations can be derived, both for full SME or a certain sector of it \cite{Colladay:2001wk,Rubtsov:2012kb}. Tree-level processes in this type of models are deeply investigated. Several processes, kinematically forbidden in LI theory, may occur in its LV extension; thresholds and cross-sections of other processes may be modified \cite{Colladay:2001wk,Rubtsov:2012kb, Coleman:1997xq,Jacobson:2002hd}. These tree-level phenomena lead to several experimental constraints on SME parameters \cite{Kostelecky:2008ts}. 

Loop-level processes in theories without LI have been also studied since \cite{Chadha:1982qq}. One of the main achievements in this area is the proof of one-loop renormalizability of QED sector of SME \cite{Kostelecky:2001jc}. In that work infinite parts of one-loop diagrams have been calculated. A study of finite radiative corrections was initiated in \cite{Cambiaso:2014eba}. The authors of \cite{Cambiaso:2014eba} have shown that these corrections may be momentum-dependent and may influence the propagation of free particles (see also \cite{Potting:2011yj}). 

%However, the issue of finite parts of one-loop diagrams  is still an open question; several attempts to investigate them and the corresponding physics have been done in \cite{Cambiaso:2014eba}. A renormalization in LI theory keep the structure of quadratic Lagrangian unchanged, only existing parameters (fields, mass and couplings) are renormalized; in LV case it fails \cite{Cambiaso:2014eba}, and some novel structures may appear in the effective Lagrangian.

The last statement can be illustrated in the following way. A chain of one-particle reducible diagrams (polarization operators or self-energy) can be summed into one-loop modified propagator. The poles of the propagator determine modified dispersion relation for the corresponding particle. In LI case the corrections to the denominator of the propagator reduce to the renormalization of fields and parameters; but in LV case not. There are two examples in standard physics: photon dispersion in external magnetic (see e.g. \cite{Kuznetsov:2013sea}) and  gravitational \cite{Hollowood:2009qz} field. Both external classical fields violate LI, and in both cases photon velocity depends on its energy, as well as on the external field parameters. The similar situation may occur if LI is broken at fundamental level like in SME.
%A test of this issue in a simplified model is a scope  of this article. 
%Loop-level processes in theories without LI have been also studied since \cite{Chadha:1982qq}, but there is still a lot of unknown in this area. It was shown that QED sector of SME is one-loop renormalizable \cite{Kostelecky:2001jc}, several aspects of renormalization have been investigated later \cite{Cambiaso:2014eba}.

Cambiaso et.al.\cite{Cambiaso:2014eba} have calculated momentum-dependent radiative correction to the electron dispersion relation in a leading order on SME parameters in a simplified CPT-even nonbirefringent version of SME. They used a general technique of calculations in SME ---  perturbative treatment of LV parameters. However, in the most simplified models, which are both CPT-even and isotropic, nonperturbative treatment can be possible. Thus, both external states and propagators may be computed exactly on LV parameters. This allows us to use these expressions to compute loop diagrams nonperturbatively at all orders in LV parameters. The aim of this article is to adopt this nonperturbative approach to the calculation of finite momentum-dependent part of the photon polarization operator and which allows us to calculate one-loop radiative correction to the photon dispersion relation. The charge of this nonperturbative treatment is the restriction to a very limited number of LV parameters.

The paper is organized as follows. In Sec.2 we describe a simplified model that we use for calculations. In Sec.3 we provide one-loop calculation of the photon polarization operator, summarize one-loop radiative corrections to the photon propagator, and compute the modified photon dispersion relation. In Sec.4 we establish bounds on LV parameters for charged fermions from photon timing observations; Sec.5 is devoted to discussion. 

\section{The Model}

The QED sector of SME is described by the Lagrangian \cite{Colladay:1998fq}
\begin{eqnarray}\label{bare-lagrangian-mSME}
\mathcal{L}_{SME} \!& = &\! -\frac{1}{4}F_{\mu\nu}F^{\mu\nu}+ i\,\overline{\psi}\,
\Gamma^\mu D_\mu\, \psi
-\overline{\psi}\, M \,{\psi}-\\
&&\! -\frac{1}{4}(k_{F})_{\mu\nu\rho\sigma}F^{\mu\nu} F^{\rho\sigma}+\frac{1}{2}(k_{AF})^{\kappa}\epsilon_{\kappa\lambda\mu\nu} A^{\lambda} F^{\mu\nu}\;,\nonumber
\end{eqnarray}
where the  constants $(k_{F})_{\mu\nu\rho\sigma}$,
$(k_{AF})^{\mu}$ control LV in photon sector, and $\Gamma^\mu$ and $M$ have the following form:
\begin{eqnarray}\label{covar-Gamma-M-B}
\Gamma^\mu\!&=&\!\gamma^\mu+c^{\mu\nu}\gamma_\nu+d^{\mu\nu}\gamma_5\gamma_\nu+if^\mu
+\frac{1}{2} g^{\lambda\nu\mu}\sigma_{\lambda\nu}+e^\mu\,,\nonumber\\
M\!&=&\!m+a^\mu\gamma_\mu+b^\mu\gamma_5\gamma_\mu+\frac{1}{2} H^{\mu\nu}\sigma_{\mu\nu}\,.
\end{eqnarray}
Here $a^\mu$, 
$b^\mu$, 
$c^{\mu\nu}$, 
$d^{\mu\nu}$, 
$e^\mu$, 
$f^\mu$, 
$g^{\lambda\mu\nu}$, and 
$H^{\mu\nu}$
are the constants controlling all type of Lorentz and CPT violation. The electromagnetic field strength $F_{\mu\nu} \equiv \partial_\mu A_\nu - \partial_\nu A_\mu $ and the covariant derivative $D_\mu = \partial_\mu - ieA_\mu$ are defined in a usual way, $e$ is the electron charge. The Greek indices $\mu,\nu...$ are raised and lowered by using the Minkowski metric. It was shown that the Lagrangian (\ref{bare-lagrangian-mSME}) is one-loop renormalizable \cite{Kostelecky:2001jc} and does not loose renormalizability in curved spacetime \cite{deBerredoPeixoto:2006wz}.

Note that the renormalization procedure \cite{Kostelecky:2001jc} implies a certain mixing of some coefficients in (\ref{covar-Gamma-M-B}).
Nevertheless, there are several separate sectors in the parametric space which are compatible with the structure of SME renormalization, for example C- and CPT-even sector in which all coefficients  except $(k_{F})_{\mu\nu\rho\sigma}$ and $c_{\mu\nu}$ are set to zero\footnote{This model, with the reduced form of $(k_{F})_{\mu\nu\rho\sigma}$, has been considered in \cite{Cambiaso:2014eba}}. In this work we restrict ourselves to SO(3)-invariant sector in which all time components of $(k_{F})_{\mu\nu\rho\sigma}$ and $c_{\mu\nu}$ are set to zero while each space component is characterized by a single parameter:
\begin{equation}\label{KFC}
(k_{F})_{\mu\nu\rho\sigma} = c_\gamma\, \cdot\, \delta_\mu^i \delta_\nu^j \delta_\rho^k \delta_\sigma^l \left( \delta_{ik}\delta_{jl} - \delta_{il}\delta_{jk}\right), \qquad c_{\mu\nu} = c_e \,\cdot\, \delta_\mu^i \delta_\nu^j\, \delta_{ij}. 
\end{equation}
The Greek indices $\mu,\nu...$ run from $0$ to $3$ while the Latin indices $i,j..$ are spacelike, taking the values $1,\,2,\,3$.
Substituting the expressions (\ref{KFC}) to the Lagrangian (\ref{bare-lagrangian-mSME}), we obtain the following model:
\begin{align}\label{L1}
\mathcal{L}_{model}=&-\frac{1}{4}F_{\mu\nu}F^{\mu\nu}+i\bar{\psi}\gamma^\mu D_\mu\psi - m\bar{\psi}\psi - \notag \\
 - &\frac{c_\gamma}{2}F_{ij}F^{ij} - i c_e\bar{\psi}\gamma^i D_i\psi. 
\end{align}
Here the parameters $c_\gamma,\;c_e$ determine the maximal velocities for photon and electron/positron respectively. Namely, the Lagrangian (\ref{L1}) yields the following tree-level dispersion relations:
\begin{align}
\gamma: \qquad & k_0^2 = \left(1+c_\gamma\right)^2\vec{k}^2 \simeq \left(1+2c_\gamma\right)\vec{k}^2, \label{PhDR} \\
e^\pm: \qquad & E^2 = \left(1+c_e\right)^2 p^2 + m^2 \simeq \left(1+2c_e\right)\vec{k}^2 + m^2. \label{EDR}
\end{align}

Let us notice that one can redefine fields and coordinates in a way that one of the parameters $(c_\e,\, c_\gamma)$ disappears from the Lagrangian (\ref{L1}), since only the difference $(c_e-c_\gamma)$ carries physical meaning\footnote{In particular, redefinitions $x_i \to (1+c_e)x_i,\, A_i\to (1+c_e)^{-1}A_i, \, \psi\to(1+c_e)^{3/2}\psi$ in the action $S=\int d^4z \mathcal{L}_{model}$ remove the parameter $c_e$ from the fermionic part of the model. In terms of these new fields, the combination $(c_\gamma - c_e)$ appears in the photon dispersion relation (\ref{PhDR}) instead of $c_\gamma$.}.
However, for completeness we will keep both of them for further analysis.

The model (\ref{L1}) has been described in \cite{Anselmi:2011ae}, where the exact expression for external states and propagators have been obtained. The electron propagator has the same form as in the standard case:
\be\label{ElectronPropagator}
S(p)=\frac{i\left(\gamma^\mu \hat{p}_\mu-m\right)}{\hat{p}^2-m^2},
\ee
where $\hat{p}_\mu$ is no longer the electron four-momentum but $\hat{p}_\mu = (p_0,(1+c_e)\,p_i)$, $\hat{p}^2 = \hat{p}_\mu\hat{p}^\mu$. 
In order to write the photon propagator, one should first add to the Lagrangian (\ref{L1}) a gauge-fixing term, which, in order to get rid of non-diagonal terms in propagator, is convenient to take in the form 
\begin{equation} \label{gf}
\mathcal{L}_{gf} \, = \, -\frac{1-2c_\gamma}{2}\left( \partial_0 A_0 - (1+2c_\gamma) \partial_i A_i\right)^2.
\end{equation}
Inverting the quadratic part of the photon Lagrangian in (\ref{L1}) with (\ref{gf}), one obtains the photon propagator (cf. \cite{Anselmi:2011ae,Rubtsov:2012kb}) in ``pseudo-Lorentz" gauge:
\begin{equation}\label{PhotonPropagator}
D^{\mu\nu} (k) = -i\frac{\mbox{diag}\left( (1+2c_\gamma), \, -1,\, -1,\, -1 \right)}{k_0^2 - (1+2c_\gamma)\vec{k}^2}.
\end{equation}
For our calculation the photon propagator in the Coulomb gauge $\partial_i A_i = 0$ is more convenient:
\begin{equation}
\label{CoulombPropagator}
D^{00} (k) = i \frac{1}{(1+2c_\gamma)\vec{k}^2}, \qquad D^{0i}(k)=0, \qquad D^{ij}(k) = i \frac{\delta^{ij} - \frac{k^ik^j}{\vec{k}^2}}{k_0^2 - (1+2c_\gamma)\vec{k}^2}.
\end{equation}
To obtain the polarization operator, one also needs to know the photon-fermion vertex, which now takes the form:
\be\label{Gamma}
\Gamma_\mu = -i e\l \gamma_0, (1+c_e)\gamma_i\r.
\ee
We apply the expressions (\ref{ElectronPropagator}, \ref{PhotonPropagator}, \ref{CoulombPropagator}, \ref{Gamma}) to compute the one-loop photon polarization operator.
%------------------------------------------------------------------%

\section{Photon polarization operator}

The aim of this section is to calculate the photon polarization operator in one-loop approximation, and subsequently resum one-loop contributions into the photon propagator.

Following rules of the standard perturbation theory, let us write the expression for the photon polarization operator in one-loop approximation:
\be\label{PIMUNU}
\Pi_{\mu\nu}(k) = \int \frac{d^4 k_l}{(2\pi)^4} \Tr \left[\Gamma_\mu S(k+k_l) \Gamma_\nu S(k_l)\right],
\ee
where $\Gamma_\mu$ and $S(k_l)$ are vertex and propagator defined in the previous section and $k_l$ is the loop momentum.  Rescaling the loop momentum: $k^i_l \to (1+c_e) k^i_l$ and introducing "hat" momentum $\hat{k}=\left(k_0, (1+c_e) k_i\right)$, we can represent the components $\Pi_{00}$, $\Pi_{0i}$ and $\Pi_{ij}$ of the photon polarization operator (\ref{PIMUNU}) via LI ones:
\be\label{Picomps}
\Pi_{00}(k) = (1+c_e)^{-3}\Pi_{00}^{LI}(\hat{k}); \qquad \Pi_{0i}(k) = (1+c_e)^{-2}\Pi_{0i}^{LI}(\hat{k}); \qquad \Pi_{ij}(k) = (1+c_e)^{-1}\Pi_{ij}^{LI}(\hat{k}).
\ee
Here $\Pi^{LI}_{\mu\nu}(k)$ is the standard LI polarization operator 
\be\label{pili}
\Pi_{\mu\nu}^{LI}(k) = \left(\eta_{\mu\nu} - \frac{k_\mu k_\nu}{k^2} \right) k^2 \Pi(k^2),
\ee
where $\Pi({k}^2)$ is expressed via dimensional regularization technique as
\be\label{PI}
\Pi(k^2)=-\frac{e^2}{2\pi^2}\int_0^1 dx\,x(1-x)\left[ \frac{1}{\eps} + \ln 4\pi - \gamma_E - \ln\frac{m^2-x(1-x){k}^2}{\mu^2}\right].
\ee
Here  $\eps=4-2d$ tends to zero as the number of dimensions $d$ tends to 4, $\gamma_E$ is the Euler constant. By using eqs. (\ref{Picomps})-(\ref{PI}), we rewrite the LV polarization operator as follows:
\be\label{PII}
\Pi_{\mu\nu}(k)=\left[ (1-c_e)k^2 (P_1)_{\mu\nu} - 2c_e \vec{k}^2 (P_2)_{\mu\nu}\right]\Pi(\hat{k}^2),
\ee
where we have introduced two projection operators
\begin{equation}\label{Proj}
P_1^{\mu\nu} = \eta^{\mu\nu} - \frac{k^\mu k^\nu}{k^2}, \qquad P_2^{\mu\nu} = -\delta_i^\mu \delta_j^\nu \left( \delta^{ij} - \frac{k^i k^j}{\vec{k}^2}\right),
\end{equation}
with the properties $
P_{1\nu}^\mu \, P_{1\lambda}^\nu = P_{1 \lambda}^\mu, \quad
P_{2\nu}^\mu \, P_{2\lambda}^\nu = P_{2 \lambda}^\mu, \quad
P_{1\nu}^\mu \, P_{2\lambda}^\nu = P_{2 \lambda}^\mu,
$
and
\be\label{PIfin}
\Pi(\hat{k}^2)=\frac{e^2}{2\pi^2}\int_0^1 dx\,x(1-x) \ln\left(1-x(1-x)\frac{\hat{k}^2}{m^2}\right) + C_\Pi,
\ee
where $C_\Pi=-\frac{e^2}{12\pi^2}\left(\frac{1}{\varepsilon} + \ln 4\pi - \gamma_E -\ln \frac{m^2}{\mu^2} \right)$.

The polarization operator (\ref{PII}) contains two infinite terms proportional to the projectors $P_1^{\mu\nu}$ and $P_2^{\mu\nu}$. In the renormalization procedure, they are contracted with their counterterms appearing from renormalizing the electromagnetic field $A_\mu$ and parameter $c_\gamma$ respectively (see \cite{Kostelecky:2001jc} for detailed calculations in SME). Dependent on the concrete subtraction scheme, renormalized constant $C_\Pi$ may take different values. Being interested in propagation of a free photon, we apply on-shell subtraction scheme. From physical grounds, we assume no radiative corrections for soft on-shell photons. This can be achieved by setting $C_\Pi=0$, which results in $\Pi(0)=0$.

Let us sum a chain of one-particle reducible diagrams into the modified photon propagator. This procedure is simpler if we take photon propagator in Coulomb gauge (\ref{CoulombPropagator}) (see Appendix A for comparison with "pseudo-Lorentz" gauge). The summation goes independently for the time and space parts, and lead to the result (cf. (\ref{CoulombPropagator})):
\begin{align}
\label{D1-loopl1}
&D^{00}_{1-loop} (k) =  \frac{i}{\vec{k}^2\left( 1+2c_\gamma -(1-c_e)\Pi(\hat{k}^2)\right)}, \qquad \qquad D^{0i}_{1-loop} (k)=0, \\
\label{D1-loopl2}
&D^{ij}_{1-loop} (k) = \frac{i}{1-\Pi(\hat{k}^2)(1-c_e)} \,\cdot\, \frac{\delta^{ij} - \frac{k^i k^j}{\vec{k}^2}}{k_0^2-\vec{k}^2(1+2c_\gamma+2(c_\gamma-c_e)\Pi(\hat{k}^2))} .
\end{align}
Time components of the propagator, $D^{00}$ and $D^{0i}$, are the same as in tree-level propagator (up to the coefficient), the space component $D^{ij}$ keeps its tensor structure proportional to the projector $P_2^{\mu\nu}$ but the pole structure of the denominator changes. It is known that the position of the pole in (\ref{D1-loopl2}) determines dispersion relation for a free photon.  The pole from the first term, $1/\left[ 1-\Pi(\hat{k}^2)(1-c_e) \right]$, is a usual Landau pole. The pole from the second term of (\ref{D1-loopl2}) is physical. Hence, to find the photon dispersion relation explicitly one should solve the equation
\begin{equation}
\label{denominator}
k_0^2-\vec{k}^2(1+2c_\gamma+2(c_\gamma-c_e)\Pi(\hat{k}^2))=0,
\end{equation}
where $\Pi(\hat{k}^2)$, given by (\ref{PIfin}), includes the zeroth component of the momentum $k_0$ as well. For this purpose we apply an iteration procedure: we start from the tree-level dispersion relation (\ref{PhDR}) at zero order and consider  $\Pi(\hat{k}^2)$ as a small perturbation suppressed by $\alpha_{em} = e^2/4\pi$. At the first order on $\alpha_{em}$, the dispersion relation is 
\be\label{DRgeneral}
k_0^2=\vec{k}^2\left(1+2c_\gamma+2(c_\gamma-c_e)\Pi_\epsilon(\vec{k}^2)\right),
\ee
where
\be\label{pi1}
\Pi_\epsilon \left(\vec{k}^2 \right)=\frac{e^2}{2\pi^2}\int_0^1 dx\,x(1-x) \ln\left(1+2\,(c_e-c_\gamma)x(1-x)\frac{\vec{k}^2}{m^2}\right).
\ee
Let us introduce a notation $y \equiv \left( c_e-c_\gamma\right)\frac{\vec{k}^2}{m^2}$ and perform integration in (\ref{pi1}) analytically. In the case $y>-2$ one obtains
\be\label{exactPi}
\Pi_\epsilon\left(\vec{k}^2\right) =\frac{e^2}{2\pi^2} \left[
\frac{y-1}{3y}\sqrt{\frac{y+2}{y}}\arcth\sqrt{\frac{y}{y+2}}+\frac{1}{3y}-\frac{5}{18}\right];
\ee
otherwise ($y<-2$) the polarization operator gains nonzero imaginary part. According to the optical theorem, the process of photon decay to an electron-positron pair $\gamma \to e^+e^-$ takes place in this case\footnote{The condition $y=-2$, which determines the position of logarithmic cut at the momentum complex plane, coincides with an energy threshold condition for the photon decay process.}. This process is extremely fast \cite{Rubtsov:2012kb}, so any phenomenological consideration of modified dispersion relation seems to be irrelevant.

Let us go back to the case $y>-2$, when the photon decay is kinematically forbidden. The expression (\ref{exactPi}) can be simplified in two limiting cases. In the limit
of large positive $y \gg 1$, one obtains
$\Pi_\epsilon=\frac{q^2}{12\pi^2}\l \ln (2 y) - \frac{5}{3}\r$ (cf. (\ref{pi1})).
Then, the photon dispersion relation obtains logarithmic correction:
\be\label{gg1}
k_0^2=\vec{k}^2\left[ 1 + 2c_\gamma + \frac{e^2}{6\pi^2}(c_\gamma - c_e)\cdot \left[\ln \l2\l c_e-c_\gamma\r\frac{\vec{k}^2}{m^2}\r - \frac{5}{3} \right] \right], \qquad \left( c_e-c_\gamma\right)\frac{\vec{k}^2}{m^2} \gg 1.
\ee
The radiative correction of photon dispersion relation results in the dependence of the photon velocity on its energy. Hence, the physical velocity, defined as $c_\gamma^{ph} \equiv \frac{\partial k_0}{\partial |\vec{k}|}$ (see (\ref{DRgeneral}), (\ref{pi1})), is no longer a constant.
%This phenomenon may be used to constrain LV, to be discussed in the next section.
 In the limit $y \gg 1$ the physical photon velocity obtains negative radiative correction:
\begin{equation}
\label{clog}
c_\gamma^{ph} = 1+c_\gamma - \frac{e^2}{6\pi^2}\cdot \left( c_e - c_\gamma\right)\cdot \ln\left(2(c_e-c_\gamma)\frac{\vec{k}^2}{m^2}\right).
\end{equation}
The expression (\ref{clog}) coincides with the result for renormalization group analysis for $c_\gamma$ obtained in  \cite{Kostelecky:2001jc} if we take the renormalization group scale $\mu$ in \cite{Kostelecky:2001jc} as  $\mu=\sqrt{c_e - c_\gamma}\,E_\gamma$. This can be explained in the following way. Let us set $c_e = 0$ (this can be achieved via field and coordinate redefinition, see the footnote after formula (\ref{EDR})). The photon polarization operator (\ref{PII}) considered on-shell  may be interpreted as an off-shell polarization operator calculated in LI theory with the squared photon momentum 
\begin{equation}\label{off-shell_analogy}
q^2 \equiv E_\gamma^2 \,-\,\vec{k}^2 = 2(c_\gamma - c_e)\,E_\gamma^2.
\end{equation}
%This value plays the role of the ``transferred momentum" --- the standard meaning of the renormalization group scale.
The case of the logarithmic correction $y \gg 1$ corresponds to $q^2 \gg m^2$.

In the opposite limit $|y|\ll 1$ the expression (\ref{DRgeneral}) can be expanded into series in $y$, the leading term is $\Pi_\epsilon = \frac{e^2}{30\pi^2}\,y + O(y^2)$. The effective photon dispersion relation (\ref{DRgeneral}) acquires an extra quartic term in the leading order:
\begin{equation}\label{ll1}
k_0^2 = \vec{k}^2\,(1+2c_\gamma) - \frac{\vec{k}^4}{M_{LV,e}^2}, \qquad \quad  \left| c_e-c_\gamma\right|\frac{\vec{k}^2}{m^2} \ll 1.
\end{equation}
Here the effective LV mass scale $M_{LV,e}$ is defined as
\begin{equation}\label{MLV}
M_{LV,e} =\frac{\sqrt{15}\,\pi}{e}\,\cdot\,\frac{m}{|c_e - c_\gamma|}.
\end{equation}
%Determining physical photon velocity in a similar to the previous case way, one obtains $c_\gamma^{ph}=1+c_\gamma -\frac{3k^2}{2M_{LV}^2}$. 
Let us note that the sign minus before the quartic term in (\ref{ll1}) appears for both positive and negative $y$. The next-to-leading term in  (\ref{ll1}) is expected to be of the order $O\left(\left( c_\gamma -c_e \right)^{-1}\,\frac{\vec{k}^6}{M_{LV,e}^4}\right)$ \footnote{However, this term is of the order of $\alpha_{em}^2$, as well as the contribution from two-loop correction to the polarization operator.}, and may take sign plus or minus depending on the sign of $(c_\gamma - c_e)$. Similarly to the previous case, physical photon velocity depends on its momentum: $c_\gamma^{ph} =1+c_\gamma -\frac{3\vec{k}^2}{2 M_{LV,e}^2}$. This dependence may be tested experimentally, which we study in the next section.

\section{Experimental constraints on LV in fermion sector from photon observations}

In the previous section we have calculated the radiative correction to the photon dispersion relation in QED, considering an electron running in the loop in the photon polarization operator. However, in the full Standard Model, the photon polarization operator in fact gets corrections not only from electrons but from all charged particles presenting in the theory. Assuming tree-level LV for a certain charged particle, one can perform a machinery similar to the aforementioned one, and obtain the radiative correction to the photon velocity caused by this particle. For two or more particles with nonzero analog of $c_e$ (electric charges are assumed to be the same), in the first order on $\alpha_{em}$ the full correction to the photon dispersion relation is the sum of corrections calculated for corresponding particles. 
%the photon polarization operator also includes contributions with muon and tau inside the loop, as well as other charged particles. Since the polarization operator include sum of the contributions linearly, the correction to the dispersion relation (\ref{DRgeneral}) include contribution of such charged spieces linearly, $\Pi_\epsilon$ in (\ref{DRgeneral}) may be replaced with sum of $\Pi_\epsilon$ over all spieces (it fails for different electric charges). Thus, hypothetically photon may obtain both logarithmic and quartic corrections from different fermions.
For two or more quartic corrections, associated with charged fermions, the effective LV mass scale (\ref{MLV}) is determined as 
\begin{equation}\label{sumMass}
M_{LV}^{-2}= \sum_f M_{LV,f}^{-2}.
\end{equation}
Here we summed over all charged fermions $f$, $M_{LV,f}$ is defined by the formula (\ref{MLV}) for a concrete charged fermion with maximal tree-level velocity $1+c_f$, electric charge $e_f$ and mass $m_f$. If $M_{LV,f}^{-2}$ for a certain fermion significantly exceeds the same parameter for other charged fermions, we can set $M_{LV}\,\simeq M_{LV,f}$ with a good accuracy. Thus, we can treat LV coefficients $c_f$ for different charged fermions separately.

Let us turn to the experimental constraints. %We have shown that LV in fermion sector, due to one-loop corrections, lead to corrections in the dispersion relation for a free photon. This leads to the energy dependence of the photon velocity, absent at tree level. This fact can be used to constrain LV in fermion sector from direct photon observations.
The best direct constraints on photon velocity are based on photon time-of-flight  analysis for fast distant astrophysical sources. Thus, in presence of LV, characterized by the dispersion relation (\ref{ll1}), high-energy photons from a source would arrive later than low-energy ones. The best constraint of this type \cite{Vasileiou:2013vra} is based on timing of GRB 090510 event, observed by FERMI-LAT \cite{Fermi-LAT}. In the analysis \cite{Vasileiou:2013vra} quartic dispersion relation (\ref{ll1}) has been tested for photon energies up to $150$ MeV,
 and the lower bound on LV mass scale $M_{LV} = M^{GRB}_{LV} \equiv 1.3\cdot 10^{11}$ GeV was established at $95\%$ CL. In other words, $M_{LV}$ calculated by formulas (\ref{MLV}, \ref{sumMass}) should exceed $M^{GRB}_{LV}$. Taking into account the expression for $M_{LV}$ for a certain fermion $f$, we obtain
\begin{equation}\label{genbound}
|c_f-c_\gamma| < \frac{\sqrt{15}\,\pi}{e_f}\cdot \frac{m_f}{M_{LV}^{GRB}} \simeq 3\cdot 10^{-10}\cdot \left(\frac{e_f}{e}\right)^{-1}\cdot\left(\frac{m_f}{\mbox{GeV}}\right).
\end{equation}
%Here we consider not only electron inside the loop but an arbitrary massive charged fermion with maximal tree-level velocity $c_f$, electric charge $e_f$ and mass $m_f$.
The formula (\ref{genbound}) is valid under the condition $|y| \ll 1$, which leads to  $|c_f-c_\gamma| \ll \left( m_f/150\,\mbox{MeV}\right)^2$. This condition, combined with (\ref{genbound}), is satisfied at least for leptons.

\paragraph{Bounds in lepton sector.} In the Table 1 we present bounds on the value $|c_f-c_\gamma|$, where $c_f$ rely on three generations of leptons. Comparison with the current bounds \cite{Kostelecky:2008ts} for each particle is also presented in Table 1. Our bound for electron is weaker than the current one \cite{Altschul:2010na}, obtained from the absence of anomaly synchrotron losses at LEP. However, for heavy leptons the situation changes. The bound for muon is of the same order as the current one; the bound on tau-lepton is one order of magnitude better.
\begin{table}[h]
\begin{center}
\begin{tabular}{|c|c|c|} 
\hline
$\,$&our bound &current bounds\\ 
\hline 
$\,$&$\,$&\\
electron & $1.5\cdot 10^{-13}$  & $10^{-15}$ \cite{Altschul:2010na}  \\
muon  & $3\cdot 10^{-11}$ & $10^{-11}$ \cite{Altschul:2006uw}\\
tau-lepton &$1.2\cdot 10^{-9}$ & $10^{-8}$ \cite{Altschul:2006uw} \\
%t-quark & $8\cdot 10^{-8}$ & $10^{-2}$ \cite{Abazov:2012iu}\\
\hline
\end{tabular}
\end{center}
\caption{Bounds on $\left| c_f-c_\gamma\right|$ for 3 generations of leptons.}
\end{table}
Since the bound on $\left| c_f-c_\gamma\right|$ for an electron is significantly better than for heavy leptons, the muon and tau bounds from Table 1 may be considered as bounds only on $|c_\mu|$ and $|c_\tau|$. We do not consider the case of fine-tuning $\left| c_e-c_\gamma\right| \ll c_\gamma$ here.

\paragraph{QCD sector. The bound on $c_f$ for top-quark.} The full photon polarization operator includes  the contribution from QCD sector as well. The following issue arises: should we work in the perturbative regime and consider quarks running in the loop, or work in the nonperturbative regime and consider effective theory? 

Following the analogy with the off-shell polarization operator (see the end of the previous section), we compare the ``transferred momentum" $q^2 \equiv E_\gamma^2-\vec{k}^2$ with $\Lambda_{QCD}^2$: at large $q^2 \gg \Lambda_{QCD}^2$ the QCD corrections are perturbative and small; in the opposite case $q^2 \ll \Lambda_{QCD}^2$ the perturbative treatment is not applicable. Using (\ref{off-shell_analogy}), let us rewrite the condition for the perturbative regime as 
 \begin{equation}\label{lqcd}
 |c_q| \gg \frac{1}{2}\frac{\Lambda_{QCD}^2}{E_\gamma^2}.
 \end{equation}
Here $c_q$ refers to the parameter $c_f$ for quarks. The energy scale of GRB bound $E_\gamma^{GRB}=150$ MeV \cite{Vasileiou:2013vra} is too small to make any bounds for quarks. Let us take another timing constraint from the flare of active galaxy PKS 2155-304 \cite{HESS:2011aa}, which is a bit weaker than the GRB bound \cite{Vasileiou:2013vra} but based on the observation of more energetic photons. The analysis of the flare performed by H.E.S.S. collaboration \cite{HESS:2011aa} set the bound  $M_{LV} > 6.4 \cdot 10^{10}$ GeV; photons with energies $E_\gamma\,\sim\, 0.25 - 4$ TeV (mean energy $1$ TeV) were considered. For these energies the condition (\ref{lqcd}) takes the numerical value
 \begin{equation}\label{lqcdnum}
  |c_{q}| \gg 2.4 \cdot 10^{-8}.
 \end{equation}
Here the value $\Lambda_{QCD}\approx 217$ MeV has been used.

For these values of $c_q$ and $E_\gamma$ the condition $|y| \ll 1$ may be valid only for top-quark. Performing analysis similar to (\ref{genbound}) with $M_{LV}^{AGN} \equiv 6.4 \cdot 10^{10}$ GeV instead of $M_{LV}^{GRB}$ and top-quark electric charge $e_{top} = 2/3 \,e$, one arrives to the following bound for the parameter $c_q$ for top-quark:
\begin{equation}\label{topbound}
|c_{top}| < 1.6\cdot 10^{-7}.
\end{equation}
The conditions (\ref{lqcdnum}), and $|y| \ll 1$, are valid for this bound. The bound (\ref{topbound}) is at 5 orders of magnitude better than the direct collider bound from Tevatron \cite{Abazov:2012iu} (see also the prospect for a collider bound for LHC \cite{Berger:2015yha}).
 
 For light quarks this analysis fails; one should consider the case $y\gg 1$. The bounds on $c_e$ for light quarks and/or mesons may be a scope of a separate work and should be compared with collider bounds \cite{Kostelecky:2016pyx,Karpikov:2016qvq}.
 
 %Let us notice, that the constraints on LV in quark sector requires assumption of small value of strong interaction constant, which seems to be valid only for top-quark  and fail for light quarks. Several collider bounds on LV for light quarks were recently appeared \cite{Kostelecky:2016pyx,Karpikov:2016qvq}. 

\section{Discussion}

We have calculated the finite momentum-dependent part of the  photon polarization operator in a simple model of LV QED in one-loop approximation, considering LV coefficients nonperturbatively.
The components of the one-loop polarization operator are rescaled to the components of the LI one due to the presence of a single particle inside the loop. 
 The modified photon propagator, obtained by the summation of one-particle reducible contributions to the photon polarization operator, has nontrivial poles which determine radiatively corrected dispersion relations. In different regimes the correction is either quartic on momentum either logarithmic; the physical velocity for free photon acquires radiative corrections in the corresponding way. The logarithmic correction to the photon velocity coincides with the result of renormalization group analysis for the corresponding coefficient, obtained in \cite{Kostelecky:2001jc} using infinite parts of one-loop diagrams. 
The reason of it is that the on-shell squared momentum for LV photon may be considered as off-shell squared momentum for LI polarization operator, which is the standard interpretation of the renormalization group  scale.

The position of the logarithmic cut in the momentum plane is shifted compared to the standard case, in accordance with the optical theorem.  This effect seems to be taken into account only for nonperturbative treatment of LV parameters and is usually missed in perturbative calculations (see \cite{Cambiaso:2014eba}).
 
Radiative corrections to the photon velocity, induced by a loop of a charged particle with tree-level LV, can be tested experimentally. The corresponding observations constrain LV for all charged fermions, the bounds for $\tau$-lepton and top-quark are the best in the literature.
In any case, no charged fermion cannot have large values of $c_e$; otherwise photon velocity would strongly depend on its energy (or the photon would decay).

%In the full SME one-loop corrections for the photon dispersion relation may be more complicated. For example, in CPT-violating case different polarizations of a photon may travel with different velocities depending on energy.

%It is worth to stress out that the photon dispersion relation acquire certain structures (like quartic on momentum), absent in the tree-level Lagrangian. So, the one-loop effective Lagrangian include quartic on space derivatives, as well as other high-order terms.

%, defined to give the same observables as the full theory, should include high-order space derivative terms, from which one-loop corrected dispersion relation can be obtained.  Hence, asymptotic external states for photons and correspondingly sums over polarizations are modified as well, which may be used for Feynman diagrams computation.

The modified dispersion relation for a free photon acquires quartic on momentum, and high-order terms, absent in tree-level dispersion relation (\ref{PhDR}). This fact may be also shown in terms of the effective Lagrangian for a photon. Integrating out charged fermions with tree-level LV, one should obtain an addition to the effective Lagrangian, which in the lowest order is expected to be equal to
\begin{equation} \label{EffLagr}
\delta \mathcal{L}_{eff} \sim -\frac{1}{4M_{LV}^2} F_{ij}\Delta F^{ij}.
\end{equation}
Here $M_{LV}$ is determined according formula (\ref{sumMass}) for LV charged fermions. We can integrate out all charged SM particles except the electrons and obtain QED effective Lagrangian which includes the dimension-six kinetic term (\ref{EffLagr}). Such high-dimensional terms in the LV QED Lagrangian have been considered in the literature (see \cite{Rubtsov:2012kb,Anselmi:2011ae,Mattingly:2008pw} for example). In these models QED Feynman rules are modified, so there are some changes in thresholds and cross-sections for several tree-level processes in QED. Astrophysically relevant examples of such processes are pair production by high-energy photon on photon background or in the Coulomb field \cite{Rubtsov:2012kb}. These reactions influence on the processes of photon propagation in extragalactic medium, and shower formation in the atmosphere. Experimental constraints on $M_{LV}$, based on the detection of TeV photons from astrophysical sources, are of the same order but a bit better than $M_{LV}^{GRB}\,$ \cite{Rubtsov:2016bea}.  Considerations of these processes may set a bit stringer bounds on the parameter $c_f$ for charged fermions, than presented in Table~1. Hypothetical experimental observation of ultra-high-energy ($\sim \, 10^{19}\,\mbox{eV}$) photons would establish significantly better constraints on $M_{LV}$\cite{Rubtsov:2013wwa}, and subsequently better constraints on $c_f$.

Radiative corrections to the physical velocity can be considered nonperturbatively on $c_e$ and $c_\gamma$ for electrons as well. Such calculation, perturbative on SME parameters, has been performed in \cite{Cambiaso:2014eba}. Nonperturbative calculation in our simplified model can be a good test of it. The radiative corrections to the electron velocity can be tested experimentally as well.  However, the corresponding constraints are expected to be worser than the constraints from photon velocity measurements. %Thus, these corrections
%Let us also notice, following lines of \cite{Cambiaso:2014eba}, that asymptotic external states for photons and electrons in LV QED are also modified due to one-loop corrections. These modifications
%lead to changes for thresholds and cross-sections of several tree-level processes in QED. Astrophysically relevant examples of such processes are  pair production by high-energy photon on photon background or in the Coulomb field \cite{Rubtsov:2012kb,Rubtsov:2016bea}. Considerations of these processes, including radiatively corrected Feynman rules, may set a bit stringer bounds on LV for charged fermions, than presented in Table 1.

%In the standard renormalization procedure of a Lorentz-invariant QFT in flat spacetime the renormalization group scale is usually considered as a transferred momentum in a scaterring process. Really, only parameters, running via renormalization group, in lorentz-invariant QFT are coupling constants; one can measure them only in scattering processes.  Lorentz invariance does not violated by loop corrections, and the dispersion relation is fixed.

%However, it is not only interpretation of renormalization group scaling. Thus, in curved spacetime renormalization the metric scaling parameter appears [Buchbinder, Odintsov, Shapiro].

\paragraph{Acknowledgements} The author is thankful to Sergey Sibiryakov, Dmitry Kirpichnikov, Dmitry Gorbunov, Emin Nugaev, Ivan Kharuk and Grigory Rubtsov for helpful discussions and comments on the draft of the paper. 
This work was supported by the RSF grant 14-22-00161.

\appendix
\section{Radiatively corrected photon propagator in \\ ``pseudo-Lorentz" gauge}

Let us summarize one-particle reducable contributions to the photon propagator in pseudo-Lorentz gauge (\ref{PhotonPropagator}). It is simpler perform field and coordinate rescaling (see the footnote after eq. (\ref{EDR})) in order to set   $c_\gamma = 0$ (otherwise a complicated resummation should be needed). The summation yields
\begin{equation}\label{ModPropagator}
\Delta_{\mu\nu}(k) = \frac{1}{1-\Pi(\hat{k}^2)(1-c_e)}\left[ \frac{(P_1-P_2)_{\mu\nu}}{k_0^2-\vec{k}^2} + \frac{(P_2)_{\mu\nu}}{k_0^2-\vec{k}^2(1-2c_e\Pi(\hat{k}^2))} \right],
\end{equation}
where projectors $P_1^{\mu\nu} ,\, P_2^{\mu\nu} $ were defined in (\ref{Proj}). The overall coefficient $1/\left[ 1-\Pi(\hat{k}^2)(1-c_e) \right]$ determines Landau pole, as previously. The position of the pole of the second term gives the photon dispersion relation. At first sight it seems that the propagator (\ref{ModPropagator}) describes more degrees of freedom because of another pole in the first term corresponding to a relativistic dispersion relation. This would contradict the fact that photons have two polarizations that, according to CPT, must propagate with the same velocity. In fact, the relativistic pole is a pure gauge artifact and disappears from gauge invariants. Indeed, consider the photon exchange amplitude between two conserved currents.
\begin{equation}
\mathcal{A} = J^\mu_1 \,\Delta_{\mu\nu}(k)\, J_2^\nu, \qquad k_\mu J^\mu_1 = k_\mu J_2^\mu =0. 
\end{equation}
A straightforward calculation yields
\begin{equation}\label{GaugeInv}
\mathcal{A} = \frac{1}{1-\Pi(\hat{k}^2)(1-c_e)}\cdot \frac{(1-2c_e\Pi(\hat{k}^2))J^0_1 J^0_2 - J_1^i J_2^i}{k_0^2-\vec{k}^2(1-2c_e\Pi(\hat{k}^2))}.
\end{equation}
We see that the spurious pole has been completely disappeared, and (\ref{GaugeInv}) gives the modified dispersion relation (\ref{DRgeneral}).

%Let us turn to the case $c_\gamma \neq 0$. The propagator is (\ref{PhotonPropagator}), and the problem rises up. The tensor structure of the propagator (\ref{PhotonPropagator})  does not decouple to the metrics and projectors $P_1, \, P_2$. Thus, summation do not save tensor structure of the propagator, and the summation procedure based on (\ref{PhotonPropagator}), fails. Thus, we use Coulomb gauge for the calculation.

\end{document}